\documentclass[aps,twocolumn,groupedaddress,showpacs]{revtex4}
\usepackage{graphicx}
\begin{document}
\bibliographystyle{apsrev}


\title{Fluctuation Effects on the Physical Properties of Cd$_2$Re$_2$O$_7$ Near 200 K}


\author{R. Jin$^1$}
\email[]{email address: jinr@ornl.gov}
\author{J. He$^{2,1}$}
\author{J.R. Thompson$^{2,1}$}
\author{M.F. Chisholm$^1$}
\author{B.C. Sales$^1$}
\author{D. Mandrus$^{1,2}$}
\affiliation{$^1$Solid State Division, Oak Ridge National
Laboratory, Oak Ridge, TN 37831}
\affiliation{$^2$Department of
Physics and Astronomy, The University of Tennessee, Knoxville, TN
37996}


\date{\today}

\begin{abstract}

Experimental investigation of the resistivity $\rho$,
susceptibility $\chi$, specific heat C$_p$, and Hall coefficient
R$_H$ of the pyrochlore Cd$_2$Re$_2$O$_7$ reveals the presence of
a continuous phase transition of uncertain origin with critical
temperature T* = 200 K.  Electron and X-ray diffraction
measurements indicate that a commensurate structural
transformation accompanies the changes in the electronic
properties, implying that lattice as well as electronic degrees of
freedom are involved in the transition. Remarkable scaling
relationships between $\rho$, $\chi$, and C$_p$ are observed,
indicating that fluctuations are crucial in the transition regime.
Both spin and ionic density fluctuation scenarios are considered.
\end{abstract}
\pacs{72.10.Di, 72.25.Rb, 72.80.-r, 72.80.Ga}

\maketitle

Pyrochlores containing second- and third-row transition metals
such as Ru, Os, and Re have attracted considerable interest lately
because they display novel collective phenomena
\cite{wang,mandrus,hanawa,jin,singh}. In the case of
Cd$_2$Os$_2$O$_7$, for example, the observed continuous
metal-insulator transition was interpreted both in terms of a
Slater transition \cite{mandrus}, and as a transition to an
excitonic insulating state \cite{singh}. Recently,
superconductivity was discovered in Cd$_2$Re$_2$O$_7$
\cite{hanawa,jin} making it the only known superconductor based on
the pyrochlore structure. Although superconductivity in
Cd$_2$Re$_2$O$_7$ continues to attract interest, the normal-state
properties of this material are also intriguing. In this Letter,
we present a study of the resistivity $\rho$, magnetic
susceptibility $\chi$, specific heat C$_p$, and Hall coefficient
R$_H$ of Cd$_2$Re$_2$O$_7$ single crystals over a wide temperature
range. We find that all quantities exhibit anomalies near a
characteristic temperature T* = 200 K, due to the onset of a
continuous phase transition. Electron diffraction results, which
show commensurate superlattice spots below 200 K, indicate that
the lattice plays an important role in the transition.
Quantitative analysis of the experimental data leads to the
discovery of remarkable scaling relationships between
d($\rho$T)/dT, d($\chi$T)/dT, and C$_p$ in the transition regime.
These scaling relationships imply that critical fluctuations
associated with the phase transition dominate both the electronic
transport and thermodynamic properties of Cd$_2$Re$_2$O$_7$ near
200 K.

The synthesis and initial characterization of the
Cd$_2$Re$_2$O$_7$ crystals used in this work was reported in Refs.
\cite{jin,he}. Shown in the main frame of Fig.\ 1a is the
temperature dependence of the dc electrical resistivity $\rho$
between 1.5 and 400 K, measured using a standard four-probe
technique. In general, $\rho$ increases with increasing
temperature, indicating the itinerant nature of the electrons.
However, it appears that $\rho$(T) behaves differently in
different temperature regimes. By plotting the data as d$\rho$/dT
vs. T in the inset of Fig.\ 1a, two peaks are clearly visible: one
occurs at T$_F$ $\sim$ 60 K, the other at T* $\sim$ 200 K.  Below
60 K, the resistivity has been found to exhibit a quadratic
temperature dependence \cite{jin}, suggesting that the ground
state of Cd$_2$Re$_2$O$_7$ is a moderately correlated Fermi
liquid.  The sharp peak at 200 K, caused by a kink in $\rho(T)$,
signifies a phase transition. Above T*, d$\rho$/dT is positive but
small, displaying a slight increase with increasing temperature.
Similar high-temperature behavior was also observed in the related
material Cd$_2$Os$_2$O$_7$ \cite{mandrus}.  It should be mentioned
that no thermal hysteresis was observed in the resistivity of
Cd$_2$Re$_2$O$_7$, consistent with a continuous phase transition
occurring at 200 K.

\begin{figure}
\includegraphics[keepaspectratio=true, totalheight = 7.5 in, width = 3.3 in]{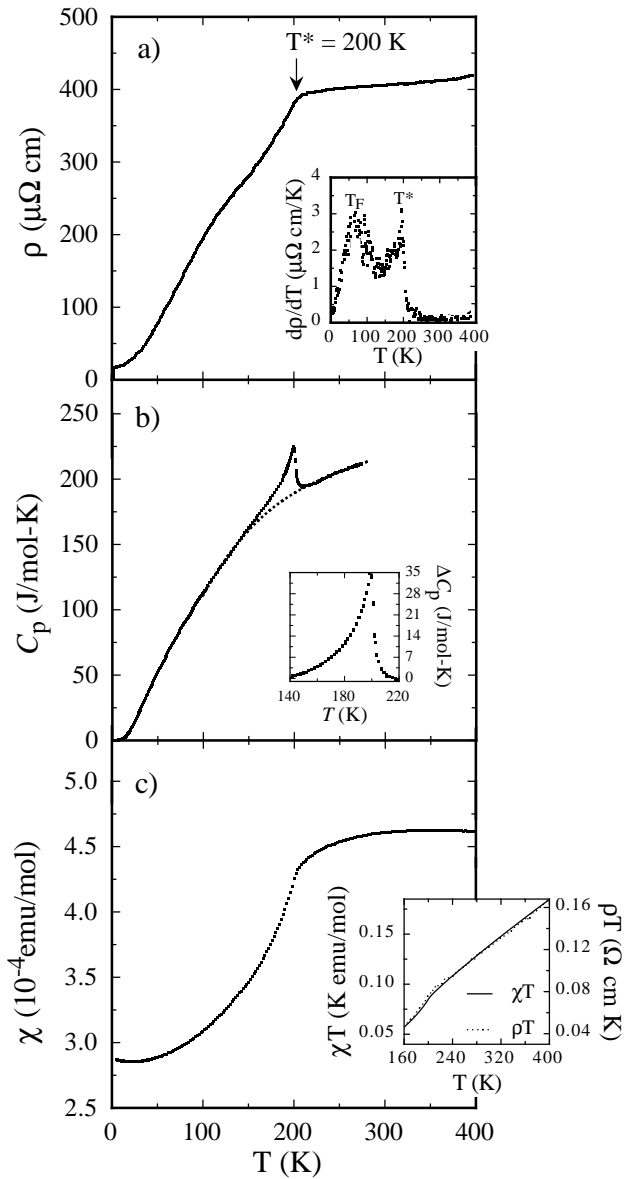}
\caption{(a) Temperature dependence of the electrical resistivity
of Cd$_2$Re$_2$O$_7$ between 1.5 K and 400 K.  The inset shows the
temperature derivative of resistivity d$\rho$/dT versus T; (b)
Specific heat versus temperature from 2 K to 280 K. The inset
shows the temperature dependence of the anomalous component of the
specific heat $\Delta$C$_p$ associated with the transition; (c)
Magnetic susceptibility of Cd$_2$Re$_2$O$_7$ obtained in an
applied field of 1 Tesla.  The inset shows the temperature
dependence of $\chi$T (solid line) and $\rho$T (dashed line)
between 160 K and 400 K.}
\end{figure}

Although the metallic behavior (d$\rho$/dT $\ge$ 0) of Cd$_2$Re$_2$O$_7$ was first reported in 1965
\cite{donohue}, the kink in resistivity was not observed in these early measurements. Quantitatively,
we obtain $\rho$(300 K) = 406 $\mu\Omega$ cm and a residual resistivity $\rho_{res}$ = 17 $\mu\Omega$
cm \cite{jin}, much lower than those given in Ref. \cite{donohue}. This suggests that scattering by
defects may smear out the resistivity anomaly at 200 K, so that this important feature becomes
visible only in samples made from very pure starting materials.

The specific heat C$_p$ of Cd$_2$Re$_2$O$_7$ was measured using a
commercial heat pulse calorimeter from Quantum Design. As can be
seen in Fig.\ 1b, C$_p$ reveals an anomaly peaked at T*.  No
latent heat or thermal hysteresis was observed in the specific
heat, consistent with a continuous phase transition.  To estimate
the magnitude and shape of the specific heat anomaly, a smooth
polynomial was fitted to the data from 100 K $\le$ T $\le$ 140 K
and 220 K $\le$ T $\le$ 280 K (see the dashed line in Fig.\ 1b).
This results in a specific heat jump of $\Delta$C$_p$(T*) = 34.0
J/mol-K (see the inset of Fig.\ 1b) and entropy S = 3.77 J/mol-K
by integrating $\Delta$C$_p$/T from 140 K to 220 K.

The dc magnetic susceptibility $\chi$ of Cd$_2$Re$_2$O$_7$ was measured using a SQUID magnetometer
from Quantum Design. Fig.\ 1c displays the temperature dependence of $\chi$ at 1 T between 2 K and
400 K. Measurements performed between 0.1 T and 2 T yield identical results. No thermal hysteresis is
noticeable within the experimental resolution. Like the resistivity, the susceptibility also reveals
a kink at T*, below which $\chi$ decreases rapidly and tends to saturate below $\sim$ 30 K.

In an itinerant electron system, a smooth increase of the magnetic susceptibility with temperature
may suggest the presence of spin fluctuations \cite{moriya}. In this case, the electron scattering is
expected to be dominated by short-range spin fluctuations as well \cite{fisher}. For
Cd$_2$Re$_2$O$_7$, the extremely weak temperature dependence of $\rho$ above the transition indicates
that the mean free path of the carriers has saturated, very likely on the order of an interatomic
spacing. To further examine the role of spin fluctuations in the electrical transport of
Cd$_2$Re$_2$O$_7$, we replot the resistivity and susceptibility as $\chi$T versus T and $\rho$T
versus T respectively in the inset of Fig.\ 1c. It is remarkable that two sets of data are coincident
between 160 and 400 K, with no adjustable parameters involved in this procedure. It is possible,
therefore, that the electrical transport, at least in the scaling region, is governed by magnetic
scattering due to spin fluctuations, with the fluctuations becoming longer ranged and slower below
the transition resulting in increased electrical conductivity.  However, as we discuss below, it is
also possible to argue that ionic density fluctuations are dominating the electronic scattering,
although in this case the scaling of the resistivity and susceptibility over a broad temperature
range becomes more difficult to understand.

The effects of spin fluctuations have been studied mainly in
systems that magnetically order. Near the magnetic critical point,
the temperature derivatives of the magnetic resistivity and
susceptibility are expected to vary like the magnetic specific
heat, $\it {i.e.}$, d$\rho$/dT, d($\chi$T)/dT, and $\Delta$C$_p$
are expected to scale as a function of temperature
\cite{fisher,fisher2}. In the case of Cd$_2$Re$_2$O$_7$ we find
that although d$\rho$/dT scales reasonably well with
$\Delta$C$_p$, a much better scaling relationship was observed by
comparing d($\rho$T)/dT to $\Delta$C$_p$.  In Fig.\ 2, we plot the
temperature dependence of d($\rho$T)/dT (solid circles),
d($\chi$T)/dT (crosses), and $\Delta$C$_p$ (open circles).
Remarkably, all curves match well between 160 K and 240 K with no
adjustable parameters except an overall normalization for each
quantity. This scaling behavior clearly demonstrates that the
anomalies observed in the specific heat, resistivity, and magnetic
susceptibility are intimately related and have a common origin.

\begin{figure}
\includegraphics[keepaspectratio=true, totalheight = 2.4 in, width = 2.4 in]{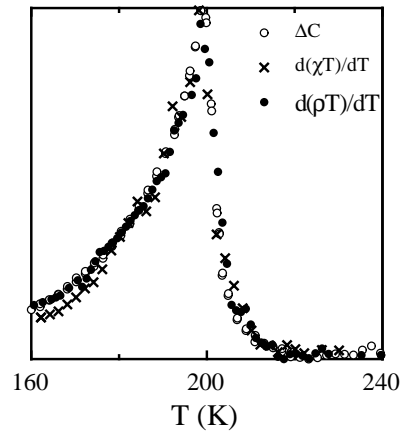}
\caption{Temperature dependence of $\Delta$C$_p$ (open circles), d($\chi$T)/dT (crosses) and
d($\rho$T)/dT (solid circles) between 160 K and 240 K.}
\end{figure}

Does the transition at 200 K involve magnetic order? The abrupt
decrease of magnetic susceptibility below 200 K is certainly
consistent with this idea, although Re compounds that display
magnetic order are rare.  In Cd$_2$Re$_2$O$_7$, the formal
oxidation state of the Re ions is 5+ with two electrons
accommodated in t$_{2g}$ manifold. Hund's rules favor parallel
spins within each manifold, and lead to spin S = 1. If the
transition is accompanied by magnetic ordering, Re ions are
expected to eliminate 2Rln(2S + 1) = 18.3 J/mol-K of entropy. This
is clearly much larger than the observed value of 3.77 J/mol-K and
suggests that the transition does not involve long-range magnetic
order, at least of localized moments.

To gain further insight into the scattering mechanisms operative in Cd$_2$Re$_2$O$_7$, Hall
measurements were performed. The Hall coefficient R$_H$ was derived from the antisymmetric part of
the transverse resistivity under magnetic field reversal at a fixed temperature. Fig.\ 3 presents the
temperature dependence of R$_H$ at 8 Tesla between 10 and 300 K. It is interesting to note that R$_H$
also reveals a cusp at 200 K, above which R$_H$ is positive and decreases with increasing
temperature. Conversely, R$_H$ decreases with decreasing temperature below 200 K and changes sign
from positive to negative near 125 K. This is consistent with band structure calculations that
predict both electron and hole sheets at the Fermi surface of Cd$_2$Re$_2$O$_7$ \cite{singh}. The
observation of the cusp-shaped R$_H$(T) suggests that the Hall contribution is also affected by the
transition at 200 K. For comparison, we replot d($\chi$T)/dT versus T in Fig.\ 3. Unlike the
resistivity, R$_H$ does not scale well with d($\chi$T)/dT above 200 K.   Nevertheless, it is worth
noting that both R$_H$ and d($\chi$T)/dT vary in the same manner below 200 K down to 60 K. This
suggests that the scattering due to fluctuations remains effective at temperatures well below the
critical point.

\begin{figure}
\includegraphics[keepaspectratio=true, totalheight = 2.4 in, width = 2.6 in]{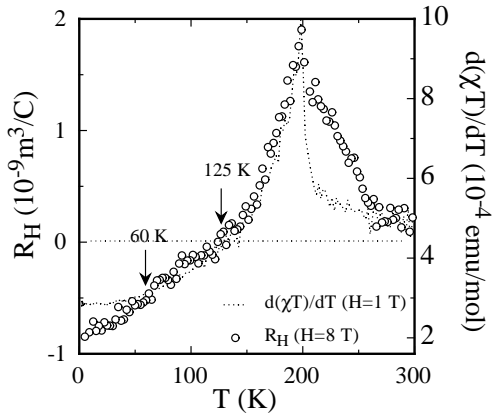}
\caption{Temperature dependence of Hall coefficient R$_H$ (open
circles). For the comparison, d($\chi$T)/dT(T) is also plotted
(dashed line).}
\end{figure}

The scaling relation between d($\chi$T)/dT and $\Delta$C$_p$ is
not unique for magnetic systems. For example, it was also observed
in the blue bronze K$_{0.3}$MoO$_3$ near the charge density wave
(CDW) transition \cite{kwok}. Theoretically, Chandra
\cite{chandra} argued that ionic density fluctuations near the
Peierls transition temperature could lead to a variation in the
density of states (DOS) at the Fermi level, and consequently
result in scaling behavior between d$\chi$/dT and $\Delta$C$_p$. A
CDW transition involves a (typically incommensurate) periodic
structural distortion, and the opening of an energy gap at the
Fermi level. For Cd$_2$Re$_2$O$_7$, both X-ray and electron
diffraction indicate that there is indeed a structural change
below T*. Fig.\ 4 presents [001] zone-axis electron diffraction
patterns obtained at 300, 200 and 108 K. At room temperature, the
Bragg spots in the diffraction pattern are consistent with the
known cubic Fd3m space group of Cd$_2$Re$_2$O$_7$. Upon cooling
from room temperature, sharp commensurate superlattice reflections
at $hkl: h,(k,l) = 2n$ become visible below 200 K, where $n$ is an
integer. These spots become more intense with decreasing
temperature. This result is also seen using X-ray powder
diffraction \cite{chako} and rotating anode measurements
\cite{gaulin}. A detailed analysis of the low temperature
structure will be presented elsewhere \cite{chako}.

In the ionic-density-fluctuation scenario, the rapid decrease of
magnetic susceptibility below 200 K results from a reduction in
the DOS at the Fermi energy and consequent reduction of the Pauli
paramagnetic contribution to the susceptibility. There are several
possibilities for the detailed nature of this sort of transition.
One possibility is that a portion of the Fermi surface develops a
gap, possibly due to a CDW instability. Band structure
calculations, however, show no evidence of Fermi surface nesting
in Cd$_2$Re$_2$O$_7$ \cite{singh,fang}.  Another possibility
involves a cooperative Jahn-Teller transition similar to that
observed in A-15 superconductors \cite{bhatt}.  In A-15's,
however, the structural transition is cubic-to-tetragonal and it
is believed that the tetragonal distortion splits a sharp peak in
the DOS into two smaller peaks with the Fermi energy residing in
the valley between the peaks.  In Cd$_2$Re$_2$O$_7$ the
low-temperature structure remains cubic so there is not an exact
parallel, but nevertheless similar physics could be driving the
transition.  Yet another possibility involves charge ordering of
the Re ions.  This possibility is supported by the fact that
Re$^{5+}$ is an uncommon oxidation state for Re.

We can estimate the fraction of states at the Fermi energy that are lost during the transition by
considering the drop in susceptibility.  The measured susceptibility is expected to contain a
paramagnetic (Pauli) contribution from spin ($\chi_{spin}$) and a core diamagnetic
contribution($\chi_{core}$). The core contribution can be obtained from standard tables with
$\chi_{core}$ = -1.72$\times$10$^{-4}$ emu/mol for Cd$_2$Re$_2$O$_7$. Thus, we estimate
$\chi_{spin}$(2K) = 4.59 $\times$10$^{-4}$ emu/mol and $\chi_{spin}$(400K) = 6.34 $\times$10$^{-4}$
emu/mol. This analysis implies that about $\sim$ 28\% of the states at the Fermi energy are lost
during the transition.

We can also evaluate the change in the DOS at the Fermi energy by considering the specific heat
anomaly.  The entropy of the itinerant electrons at 200 K is $\gamma$ $\Delta$T = (29.6
mJ/mol-K$^2$)(200 K) = 5.92 J/mol-K.  The entropy eliminated during the transition is 3.77 J/mol-K.
If we assume that the specific heat anomaly is purely electronic, we find that approximately
3.77/(5.92+3.77) or 39\% of the states at the Fermi energy are lost. This fraction is higher than
that estimated from the magnetic susceptibility. The discrepancy may be due to the assumption
S$_{el}$ = S, which is not fulfilled if the lattice contribution is significant.

\begin{figure}
\includegraphics[keepaspectratio=true, totalheight = 2.5 in, width = 3.5 in]{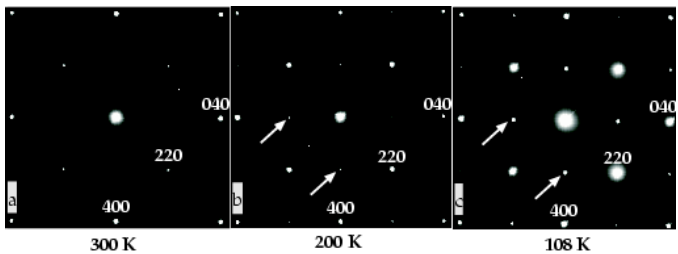}
\caption{[001] zone-axis electron diffraction patterns of Cd$_2$Re$_2$O$_7$ at (a) 300 K, (b) 200 K
and (c) 108 K. The arrows indicate the (2,0,0) superlattice spots as discussed in the text.}
\end{figure}

While the variation of the susceptibility and specific heat is
qualitatively consistent with the picture of a reduced DOS or
partially gapped Fermi-surface, it is difficult to explain why a
reduction in the DOS would result in the increased conductivity
observed below the transition. Although better metallic behavior
has been previously observed in the presence of gap due to a CDW
transition \cite{fleming}, a negative d$\rho$/dT was observed in
the critical regime. This feature is not seen in
Cd$_2$Re$_2$O$_7$. One possibility is that the removed portion of
Fermi surface is unimportant for the electrical transport.
According to Rice and Scott \cite{rice}, if the portion of Fermi
surface removed by the transition is the saddle point where the
Fermi velocity is very small and the DOS is high, the saddle
points act as scattering sinks in high-temperature phase, and
their removal can increase the conductivity. However, this theory
was developed for layered materials with 2D character and its
application to a 3D system such as Cd$_2$Re$_2$O$_7$ may be
problematic. In any case, our Hall measurements indicate that the
electronic structure has been affected by the transition, which
leads to the sign change of R$_H$.

In summary, the normal state of Cd$_2$Re$_2$O$_7$ exhibits many
intriguing properties. Thermodynamic and transport measurements
indicate that a continuous phase transition of uncertain origin
occurs at T* $\sim$ 200 K. Both X-ray and electron diffraction
measurements reveal commensurate superlattice reflections below
T*, indicating that lattice degrees of freedom play an important
role in the transition. Remarkable scaling relationships exist
between $\rho$, $\chi$, and C$_p$.  These scaling relationships
imply that critical fluctuations associated with the phase
transition dominate both the electronic transport and
thermodynamic properties of Cd$_2$Re$_2$O$_7$ near 200 K.
Scenarios involving both spin and ionic density fluctuations are
considered in the evaluation of the data. Although the observed
structural transition suggests that ionic density fluctuations are
likely to be important, a scaling relationship between $\rho$ and
$\chi$ over a wide temperature range suggests that spin
fluctuations may be playing a role as well.


\begin{acknowledgments}
We would like to thank E. W. Plummer, B.C. Chakoumakos, Z. Fang, M.D. Lumsden, S.E. Nagler, G.M.
Stocks, and L.M. Woods for helpful discussions. Oak Ridge National laboratory is managed by
UT-Battelle, LLC, for the U.S. Department of Energy under contract DE-AC05-00OR22725.  This work was
partially supported by NSF DMR 0072998.
\end{acknowledgments}

\bibliography{Cd2Re2O7IIbib.tex}

%
%

%
%

\end{document}